\documentclass[preprint,12pt]{elsarticle}

\usepackage{amsmath,amssymb}
\usepackage{amsthm,bm,arydshln}

\newtheorem{thm}{Theorem}

\newtheorem{proposition}[thm]{Proposition}
\newtheorem{corollary}[thm]{Corollary}
\newdefinition{definition}{Definition}
\newdefinition{rmk}{Remark}
\newproof{pf}{Proof}

\begin{document}

\begin{frontmatter}



\title{Laurent skew orthogonal polynomials and related symplectic matrices}

\author{Hiroshi Miki}
\address{Meteorological College, Asahi-Cho, Kashiwa 277 0852, Japan}



\begin{abstract}
Particular class of skew orthogonal polynomials are introduced and investigated, which possess Laurent symmetry.
They are also shown to appear as eigenfunctions of symplectic generalized eigenvalue problems. Furthermore, the modification of these polynomials gives some symplectic eigenvalue problem and the corresponding symplectic matrix is equivalent to butterfly matrix, which is a canonical form of symplectic matrices.
\end{abstract}


\end{frontmatter}


\section{Introduction}

Orthogonal polynomials (OPs) obey and are characterized by local recurrence relations.
It means that they are eigenvectors of structured matrices. Such matrices appear in a variety of fields and they have been studied extensively.
For example, OPs are related with tridiagonal matrices which are called Jacobi matrices and such matrices appear in many fields including numerical algorithm, stochastic process and so on.
Hence OPs play important roles in solving these problems \cite{Karlin,Papageorgiou,Vinet}.

In this paper, we focus on symplectic matrices, which appear in solving the algebraic Riccati equation in control theory \cite{Lancaster}. With respect to the symplectic matrices, a canonical form exists and it is called (unreduced) butterfly form or butterfly matrix \cite{Banse,Benner}, which is of the form
\begin{equation}\label{buttefly_matrix}
B={\small
\left(
\begin{array}{c:c}
\begin{matrix}
b_1 & & & \\
         & b_2 &  &\\
         &          & \ddots & \\
         &           &       & b_N
\end{matrix} & \begin{matrix}
b_1c_1-a_1^{-1} & b_1d_2& &\\
b_2d_2  & b_2c_2-a_2^{-1} & \ddots  &\\
         &       \ddots   & \ddots & b_{N-1}d_N\\
         &           &     b_Nd_N & b_Nc_N-a_N^{-1}
\end{matrix}  \\ \hdashline
\begin{matrix}
a_1 & & &\\
         & a_2 &  &\\
         &          & \ddots & \\
         &           &       & a_n
\end{matrix} 
  & \begin{matrix}
a_1c_1 & a_1d_2 & &\\
   a_2d_2   & a_2c_2 & \ddots  &\\
         &       \ddots    & \ddots & a_{N-1}d_N \\
         &           &    a_Nd_N   & a_Nc_N
\end{matrix}
\end{array}
\right).}
\end{equation} 
The shape of this matrix is known to be invariant under the similarity transformation using symplectic matrices. Therefore, the eigenvectors have particular structures although their properties, especially the relationship to variants of OPs, have not been discussed yet.\par 
The aim of this paper is to find generalized OPs which relate with the symplectic matrix, especially butterfly matrix. To achieve that end, we pick up skew orthogonal polynomials (SOPs).
SOPs are originally introduced in the context of random matrix theory \cite{Dyson,Mehta}.
It is also reported that SOPs (including their variants) are related with several integrable systems \cite{Adler2,Chang,Miki}.
Unlike OPs, the basic properties of SOPs have not been discussed well and this is because SOPs do not hold local recurrence relations in general. 
However, particular SOPs are known to be related with OPs and hold local relations \cite{Adler1,Forrester2,Kodama}. Especially in \cite{Kodama}, they are shown to be related with symplectic Lie algebra.
\par
This paper is organized as follows.
In section 2, the definition and basic facts about SOPs are given and the properties of SOPs introduced in \cite{Kodama} are explained.
In section 3, new class of SOPs with local recurrence relations is introduced and they are shown to be related with OPs.
In section 4, the finite analogue of the section 3 is considered and its relationship to symplectic matrix pencil is discussed.
In section 5, some modification of the new SOPs are considered and they are shown to relate with the canonical symplectic matrices. Concluding remarks will follow.

\section{Skew orthogonal polynomials (SOPs)}
We begin with a brief review of SOPs. Let us consider the bilinear 2-form $\left< \cdot | \cdot \right>:\mathbb{C}[z]\times \mathbb{C}[z]\to \mathbb{C}$ with the skew symmetry:
\begin{equation}
\left< f(z)|g(z)\right> = -\left< g(z)| f(z)\right>,
\end{equation}
which we call skew inner product.
Skew inner product appears in random matrix theory and it is usually of the form
\begin{equation}
\left< f(x)|g(x)\right> = \iint_{\mathbb{R}^2} \mathrm{sgn}(x-y)f(x)g(y)w(x)w(y)dxdy 
\end{equation}
or
\begin{equation}
\left< f(x)|g(x)\right> = \int_{\mathbb{R}} (f'(x)g(x)-f(x)g'(x))w(x)dx.
\end{equation}
Like orthogonal polynomials defined by a given inner product, a skew symmetric inner product defines SOPs.
\begin{definition}
Given a skew-symmetric inner product $\left< \cdot | \cdot \right>$, SOPs $\{ q_n(z)\}_{n=0}^{\infty }$ are defined as a polynomial sequence which satisfies the following skew-orthogonality relation:
\begin{align}\label{def:sops}
\begin{split}
&\left< q_{2m}(z)|q_{2n+1}(z)\right> = \kappa _n\delta_{mn}\quad (\exists \kappa _n\ne 0),\\
&\left< q_{2m}(z)|q_{2n}(z)\right> = \left< q_{2m+1}(z)|q_{2n+1}(z)\right>=0,\\
&\deg (q_{n}(z))=n,\quad (m,n\in \mathbb{Z}_{\ge 0}).
\end{split}
\end{align}
\end{definition} 
\begin{rmk} SOPs of even degree $\{ q_{2n}(z)\}_{n=0}^{\infty }$ are uniquely determined up to multiple constant. Monic SOPs of even degree are hence uniquely determined. However, the following map
\begin{equation}\label{freedom1}
q_{2n+1}(z)\mapsto q_{2n+1}(z)+\lambda_nq_{2n}(z),\quad (\forall \lambda_n\in \mathbb{C})
\end{equation}
does not change the skew-orthogonality relation \eqref{def:sops} and hence SOPs of odd degree are not uniquely defined even though they are monic.
\end{rmk}
In order to avoid the above ambiguity and define SOPs uniquely, we assume SOPs be monic and the coefficient of $2n$-th degree of $q_{2n+1}(z)$ be zero, i.e.
\begin{equation}
q_{2n+1}(z)=z^{2n+1}+O(z^{2n-1})
\end{equation} 
 unless otherwise specified. 
Applying Gram-Schmidt skew orthogonalization process to the monomial basis $\{ 1,z,z^2,\ldots \}$, we obtain SOPs and their Pfaffian expression \cite{Adler2}.  

\begin{proposition}
Monic SOPs $\{ q_n(z)\}$ with respect to $\left< \cdot |\cdot \right>$ are given as follows:
\begin{align}\label{expr:pfaff}
q_{2n}(z)=\frac{\tau_{n,z}}{\tau_n},\quad q_{2n+1}(z)=\frac{\sigma_{n,z}}{\tau_n},
\end{align}
where 
\begin{align}
\begin{split}
\tau_n&=\mathrm{Pf}(0,1,\ldots ,2n-1),\\
\tau_{n,z}&=\mathrm{Pf}(0,1,\ldots ,2n,z),\\
\sigma_{n,z}&=\mathrm{Pf}(0,1,\ldots,2n-1,2n+1,z)
\end{split}
\end{align}
and entries of Pfaffians are defined by
\begin{align}
\mathrm{Pf}(i,z)=z^i,\quad \mathrm{Pf}(i,j)=\left< z^i | z^j\right>\quad (i,j\in \mathbb{Z}_{\ge 0}).
\end{align}
\end{proposition}
From the Pfaffian expression, it is straightforward to see SOPs exist iff $\tau_n\ne 0$ for all non-negative integer $n$ and we hereafter assume this condition.
This expression also allows us to derive several properties of SOPs including their spectral transformation \cite{Miki}.\par
One of the most important properties of OPs is the three term recurrence relation although such relation does not hold for generic SOPs.
However, particular class of SOPs exist that relate with OPs and hold recurrence relations \cite{Adler1,Kodama}. Especially in \cite{Kodama}, the skew inner product 
\begin{equation}\label{inner:kodama}
\left< f(z)|g(z)\right>_{K} = \int_C (f(z)g(-z)-f(-z)g(z))w(z)dz
\end{equation}
was considered and the corresponding SOPs $\{ q_n^K(z)\}_{n=0}^{\infty }$ are given as follows:
\begin{align}\label{kodama1}
\begin{split}
q_{2n}^K(z)&=p_n^K(z^2),\\
q_{2n+1}^K&=zp_n^K(z^2),\quad n=0,1,\ldots,
\end{split}
\end{align}
where $\{ p_n^K(z^2)\}$ are monic OPs with the following orthogonality relation:
\begin{equation}
\int_C p_n^K(z^2)p_m^K(z^2)zw(z)dz=0\quad (m\ne n).
\end{equation}
Since $\{ p_n^K(z^2)\}$ are OPS, they hold the three term recurrence relation
\begin{equation}
zp_n^K(z)=p_{n+1}^K(z)+a_n p_n^K(z)+b_np_{n-1}^K(z),\quad (\exists a_n,b_n\in \mathbb{C})
\end{equation}
which amounts to the following recurrence relations for $\{ q_n^K(z)\}_{n=0}^{\infty }$:
\begin{align}
\begin{split}
zq_{2n}^K(z)&=q_{2n+1}^K(z),\\
zq_{2n+1}^K(z)&=q_{2n+2}^K(z)+a_nq_{2n}^K(z)+b_nq_{2n-2}^K(z).
\end{split}
\end{align}
\section{Laurent SOPs}\label{sec:LSOPs}
In this section, we shall introduce a new skew inner product $\left< \cdot | \cdot \right>:\mathbb{C}[z,z^{-1}]\times \mathbb{C}[z,z^{-1}]\to \mathbb{C}$ that is of the form
\begin{equation}\label{lskew:inner}
\left< f(z)|g(z)\right>_L= \int_C \left( f(z^{-1})g(z)-f(z)g(z^{-1})\right) w(z)dz,
\end{equation}
where $C \subset \mathbb{C}$ is a complex domain. It is straightforward to see the skew inner product \eqref{lskew:inner} has a symmetry as follows:
\begin{equation}
\left< f(z)|g(z)\right>_L=\left< g(z^{-1})|f(z^{-1})\right>_L.
\end{equation}
Therefore we call the skew inner product \eqref{lskew:inner} as skew inner product with Laurent symmetry and also the corresponding SOPs $\{ q_n^L(z)\}_{n=0}^{\infty }$ as Laurent skew orthogonal polynomials (LSOPs). The (skew-)moments are given by 
\begin{align}\label{moments_uc}
\begin{split}
\mathrm{Pf}(i,j)=\left< z^i|z^j\right>_L =\int_C \left( z^{i-j}-z^{j-i}\right)w(z)dz. 
\end{split}
\end{align}
It should be remarked here that in case $C$ is an unit circle, such Pfaffian elements appear in random matrix theory and symmetrized random growth model \cite[eq.(10.139)]{Forrester1}.
In addition, since 
\begin{equation}\label{lskew:sym}
\left< z^{i+k}|z^{j+k}\right>_L= \left< z^i|z^j\right>_L,\quad i,j,k\in \mathbb{Z},
\end{equation}
we can immediately obtain the following ``shift formula'':
\begin{align}\label{lsym}
\begin{split}
&\mathrm{Pf}(i_1+k,i_2+k,\ldots ,i_{2n}+k)=\mathrm{Pf}(i_1,i_2,\ldots ,i_{2n}),\\
&\mathrm{Pf}(i_1+k,i_2+k,\ldots ,i_{2n-1}+k,z)=z^k\mathrm{Pf}(i_1,i_2,\ldots ,i_{2n-1},z)
\end{split}
\end{align}
for $i_i,i_2,\ldots ,i_{2n},k\in \mathbb{Z}$.
Using this shift formula and the Pfaffian expression \eqref{expr:pfaff}, we find the recurrence relations for $\{ q_n^L(z)\}_{n=0}^{\infty }$ \cite{Tsujimoto}.
\begin{proposition}
Monic SOPs $\{ q_n^L(z)\}_{n=0}^{\infty }$ hold the following recurrence relation.
\begin{align}\label{lskew:rec}
\begin{split}
z(q_{2n+1}^L(z)-\alpha_{n+1}q_{2n}^L(z))&=q_{2n+2}^L(z)-q_{2n}^{L}(z),\\
z(q_{2n}^L(z)-\beta_nq_{2n-2}^L(z))&=q_{2n+1}^L(z)-\alpha_nq_{2n}^L(z),
\end{split}
\end{align}
where
\begin{equation}\label{coeff:lskew}
\alpha_n =\frac{\sigma_n}{\tau_n},\quad \beta_n=\frac{\tau_{n+1}\tau_{n-1}}{\tau_n^2},\quad \sigma_n=\mathrm{Pf}(0,1,\ldots ,2n-2,2n). 
\end{equation}
\end{proposition}
\begin{pf}
Using \eqref{pfid1} and \eqref{pfid2}, we see
\begin{align}
\begin{split}
&\mathrm{Pf}(1,2,\ldots ,2n,0,2n+1,2n+2,z)\mathrm{Pf}(1,2,\ldots ,2n)\\
&=\mathrm{Pf}(1,2,\ldots ,2n,0,2n+1)\mathrm{Pf}(1,2,\ldots ,2n,2n+2,z)\\
&-\mathrm{Pf}(1,2,\ldots ,2n,0,2n+2)\mathrm{Pf}(1,2,\ldots ,2n,2n+1,z)\\
&+\mathrm{Pf}(1,2,\ldots ,2n,0,z)\mathrm{Pf}(1,2,\ldots 2n,2n+1,2n+2),\\
&\mathrm{Pf}(1,2,\ldots ,2n-1,0,2n,z)\mathrm{Pf}(1,2,\ldots ,2n-1,2n+1)\\
&=\mathrm{Pf}(1,2,\ldots ,2n-1,0,2n,2n+1)\mathrm{Pf}(1,2,\ldots ,2n-1,z)\\
&-\mathrm{Pf}(1,2,\ldots ,2n-1,0,2n+1,z)\mathrm{Pf}(1,2,\ldots ,2n-1,2n)\\
&+\mathrm{Pf}(1,2,\ldots ,2n-1,2n,2n+1,z)\mathrm{Pf}(1,2,\ldots 2n-1,0).
\end{split}
\end{align} 
From \eqref{lsym} and \eqref{pfaff:anti2} these relations amount to
\begin{align}\label{lskew:relation1}
\begin{split}
\tau_n\tau_{n+1,z}&=z\tau_{n+1}\sigma_{n,z}-z\sigma_{n+1}\tau_{n,z}+\tau_{n+1}\tau_{n,z},\\
\sigma_n\tau_{n,z}&=z\tau_{n+1}\tau_{n-1,z}+\tau_{n}\sigma_{n,z}-z\tau_{n}\tau_{n,z},
\end{split}
\end{align}
from which the recurrence relations \eqref{lskew:rec} are straightforwardly obtained by applying \eqref{expr:pfaff}.
\end{pf}
Deleting the LSOPs of odd degree in \eqref{lskew:rec}, we find the following recurrence formula for the LSOPs of even degree:
\begin{equation}\label{lskew:rec_even}
(z^2+1)q_{2n}^L(z)=q_{2n+2}^L(z)+z(\alpha_{n+1}-\alpha_n)q_{2n}^L(z)+\beta_n q_{2n-2}^L(z).
\end{equation}
Consider Laurent polynomial series $R_n^L(z)=z^{-n}q_{2n}^L(z)$. The relation \eqref{lskew:rec_even} then becomes
\begin{equation}\label{lskew:3-term}
\left( z+ z^{-1}\right) R_{n}^L(z)=R_{n+1}^L(z)+\left( \alpha_{n+1}-\alpha_n\right) R_{n}^L(z)+\beta_n R_{n-1}^L(z).
\end{equation} 
We immediately find from \eqref{lskew:3-term} that $R_n^L(z)$ is a polynomial of degree $n$ in $w =z+z^{-1}$ and we write $R_n^L(z)=\bar{R}_n^L(w)$.
Furthermore, since \eqref{lskew:3-term} is a three term recurrence relation, Favard's theorem in the theory of OPs \cite{Chihara} claims that $\{ \bar{R}_n^L(w)\}_{n=0}^{\infty }$ are OPs, i.e. there exists some linear functional $\mathcal{L}$ such that
\[
\mathcal{L}[\bar{R}_n^L(w)\bar{R}_m^L(w)]=h_n\delta_{mn}\quad (\exists h_n\ne 0).
\] 
The explicit form of linear functional $\mathcal{L}$ is obtained from the skew orthogonality relation \eqref{def:sops}. Using \eqref{lskew:sym}, we have for $m\ge n$
\begin{align}
\begin{split}
r_n \delta_{mn}&=\left< q_{2m}^L(z)|z^{m-n+1}q_{2n}^L(z) \right>_L \\
&= \left< z^{-m}q_{2m}^L(z)|z^{-n+1}q_{2n}^L(z)\right>_L\\
&=\left< R_m^L(z)|zR_n^L(z)\right>_L\\
&=\int_C \bar{R}_m^L(z+z^{-1})\bar{R}_n^L(z+z^{-1})(z-z^{-1})w(z)dz.
\end{split}
\end{align}
To conclude, we obtain the following theorem.
\begin{thm}
The Laurent polynomials $\{ \bar{R}_n^L(w)\}_{n=0}^{\infty }$ defined by
\begin{equation}
\bar{R}_n^L(w)=z^{-n}q_{2n}^L(z),\quad w=z+z^{-1}
\end{equation}
are monic OPs with respect to the following linear functional:
\begin{equation}\label{lskew:ops-linear}
\mathcal{L}[f(w)]=\int_C f(z+z^{-1})(z-z^{-1})w(z)dz.
\end{equation}
\end{thm}
Since  $\{ \bar{R}_n^L(w)\}_{n=0}^{\infty }$ are monic OPs, they can be expressed in terms of determinants:
\begin{equation}
\bar{R}_n^L(w)=\begin{vmatrix}
c_0 & c_1 & \cdots & c_{n-1 } & 1 \\
c_1 & c_2 & \cdots & c_{n} & w\\
\vdots & \vdots & \ddots & \vdots & \vdots \\
c_n & c_{n+1} & \cdots & c_{2n-1} & w^n 
\end{vmatrix}/ \begin{vmatrix}
c_0 & c_1 & \cdots & c_{n-1 }  \\
c_1 & c_2 & \cdots & c_{n} \\
\vdots & \vdots & \ddots & \vdots  \\
c_{n-1} & c_{n} & \cdots & c_{2n-2}
\end{vmatrix},
\end{equation}
where $c_i=\mathcal{L}[w^i]$. 
Recalling the Pfaffian expression of SOPs \eqref{expr:pfaff}, we find the non-trivial relationship between Pfaffians and determinants.
 
\begin{corollary}
The following identities hold:
\begin{align}\label{pfaff-det}
\begin{split}
&\bar{\tau}_n\equiv  \begin{vmatrix}
c_0 & c_1 & \cdots & c_{n-1 }  \\
c_1 & c_2 & \cdots & c_{n} \\
\vdots & \vdots & \ddots & \vdots  \\
c_{n-1} & c_{n} & \cdots & c_{2n-2}
\end{vmatrix}=\left. \begin{matrix}
| \mu_1 & \mu_2 & \cdots & \mu_{2n-1}\\
                       & \mu_1 & \cdots & \mu_{2n-2}\\
                       &                      & \ddots & \vdots \\
                       &                      &        & \mu_1 
\end{matrix}\right|,\\
&\bar{\sigma}_n\equiv \begin{vmatrix}
c_0 & c_1 & \cdots & c_{n-2 } & c_n \\
c_1 & c_2 & \cdots & c_{n-1} & c_{n+1}\\
\vdots & \vdots & \ddots & \vdots  & \vdots \\
c_{n-1} & c_{n} & \cdots & c_{2n-3} & c_{2n-1}
\end{vmatrix}=\left. \begin{matrix}
| \mu_1 & \mu_2 & \cdots & \mu_{2n-2} & \mu_{2n}\\
                       & \mu_1 & \cdots & \mu_{2n-3} & \mu_{2n-1}\\
                       &                      & \ddots & \vdots & \vdots \\
                       &                      &        & \mu_1 & \mu_3 \\
                       &                      &        &       & \mu_{2}
\end{matrix}\right|,
\end{split}
\end{align}
iff 
\begin{equation}\label{rel:moment}
\mu_n=\sum_{k=0}^{\lfloor \frac{n-1}{2}\rfloor}(-1)^k\binom{n-1-k}{k}c_{n-1-2k}.
\end{equation}
\end{corollary}
\begin{pf}
Since $\{ \bar{R}_n^L(w)\}_{n=0}^{\infty }$ are OPs, the coefficients of the three term recurrence relation \eqref{lskew:3-term} is given by the ratio of determinants \cite{Chihara}:
\begin{align}
\alpha_{n+1}-\alpha_n=\frac{\bar{\sigma}_{n+1}}{\bar{\tau}_{n+1}}-\frac{\bar{\sigma}_n}{\bar{\tau}_n},\quad \beta_n=\frac{\bar{\tau}_{n+1}\bar{\tau}_{n-1}}{\bar{\tau}_n^2}.
\end{align}
Therefore, we see from \eqref{coeff:lskew}
$\tau_n =\bar{\tau}_n$ and $\sigma_n=\bar{\sigma}_n$ by taking $\tau_0=\bar{\tau}_0=1,\sigma_0=\bar{\sigma}_0=0$.
With respect to \eqref{rel:moment}, the relation \eqref{lskew:inner} shows
\begin{align}
\begin{split}
\mu_n&=\mathrm{Pf}(0,n)=\left< 1|z^n\right>_L=\int_C \left( z^n-\frac{1}{z^n}\right) w(z)dz\\
&=\int_C T_{n-1}\left( \frac{z+z^{-1}}{2}\right)(z-z^{-1})w(z)dz,
\end{split}
\end{align}
where $T_n(w)$ is the Chebyshev polynomials of the second kind defined by
\begin{equation}
T_{n}\left( w\right)=\frac{\sin ((n+1)\cos^{-1}w)}{\sin (\cos ^{-1}w)}=\sum_{k=0}^{\lfloor \frac{n}{2}\rfloor}(-1)^k\binom{n-k}{k}(2w)^{n-2k}.
\end{equation}
Therefore we see from \eqref{lskew:inner}
\begin{align}
\begin{split}
\mu_n&=\mathcal{L}\left[ T_{n-1}\left( \frac{w}{2}\right) \right]=\mathcal{L}\left[\sum_{k=0}^{\lfloor \frac{n-1}{2}\rfloor}(-1)^k\binom{n-1-k}{k}w^{n-1-2k}\right]\\
&=\sum_{k=0}^{\lfloor \frac{n-1}{2}\rfloor}(-1)^k\binom{n-1-k}{k}c_{n-1-2k}.
\end{split}
\end{align}
This completes the proof.
\end{pf}

It should be remarked here that the first equation in \eqref{pfaff-det} is the same as one in \cite[Prop.2.3]{Stembridge} although the proof is different.
\section{Finite Laurent SOPs and symplectic matrix pencil}
It is well known that semi-infinite Jacobi matrix
\begin{equation}
\begin{pmatrix}
a_0 & 1   &     &\\
b_1 & a_1 & 1  &\\
    & b_2 & a_2 & 1 \\
    &     & \ddots & \ddots & \ddots 
\end{pmatrix}
\end{equation}
can be diagonalized by (monic) OPs. 
We now assume that the weight function has finite support, i.e. consider the following linear functional
\begin{equation}\label{ops:finite}
\mathcal{L}[f(z)]=\sum_{k=1}^N f(z_k)w_k.
\end{equation}
The corresponding Jacobi matrix is reduced to finite matrix:
\begin{equation}
\begin{pmatrix}
a_0 & 1 & & \\
b_1 & a_1 & 1& \\
    & \ddots & \ddots & \ddots \\
    &        & b_{N-2} & a_{N-2} & 1\\
    &        &         & b_{N-1} & a_{N-1}
\end{pmatrix}.
\end{equation}
\begin{rmk}
When we assume the linear functional $\mathcal{L}$ is positive definite and consider orthonormal polynomials $\{ \tilde{p}_n(z)\}_{n=0}^{\infty }$ instead of monic ones:
\begin{equation}
\mathcal{L}[\tilde{p}_n(z)\tilde{p}_m(z)]=\sum_{k=1}^N \tilde{p}_m(z_k)\tilde
{p}_n(z_k)w_k=\delta_{mn}\quad (w_k>0),
\end{equation}
the corresponding Jacobi matrix becomes symmetric:
\begin{equation}
\begin{pmatrix}
a_0 & \sqrt{b_1} & & \\
\sqrt{b_1} & a_1 & \sqrt{b_{2}}& \\
    & \ddots & \ddots & \ddots \\
    &        & \sqrt{b_{N-2}} & a_{N-2} & \sqrt{b_{N-1}}\\
    &        &         & \sqrt{b_{N-1}} & a_{N-1}
\end{pmatrix}.
\end{equation}
\end{rmk}
Tridiagonal matrices appear in various areas and OPs thus play an important role in analysing and solving related problems.
Here we will investigate a class of matrices related to LSOPs. In what follows, we restrict the skew inner product \eqref{lskew:inner} to finite dimension and the corresponding LSOPs are orthonormal. In other words, we shall consider the discrete LSOPs $\{ \tilde{q}_n(z)\}_{n=0}^{2N-1}$ (we omit the superscript $L$ for simplicity) satisfying 
\begin{align}\label{lsops:finite}
\begin{split}
&\left< \tilde{q}_{2m}(z)|\tilde{q}_{2n+1}(z)\right>_L = \sum_{k=1}^N (\tilde{q}_{2m}(z_k^{-1})\tilde{q}_{2n+1}(z_k)-\tilde{q}_{2m}(z_k)\tilde{q}_{2n+1}(z_k^{-1}))\tilde{w}_k =\delta_{mn},\\
&\left< \tilde{q}_{2m}(z)|\tilde{q}_{2n}(z)\right>_L=\left< \tilde{q}_{2m+1}(z)|\tilde{q}_{2n+1}(z)\right>_L=0.
\end{split}
\end{align}
We also assume $\tau_n>0$ for $n=0,1,\ldots, N$, which is equivalent to $\beta_n>0$ (and $\alpha_n\in \mathbb{R}$). Then, for $n=0,1,\ldots ,N-1$ and $z=z_1,z_2,\ldots ,z_{N}$, the LSOPs $\{ \tilde{q}_n(z)\}_{n=0}^{2N-1}$ are given by
\begin{equation}
\tilde{q}_{2n}(z)=\sqrt{\frac{\tau_n}{\tau_{n+1}}}q_{2n}^L(z),\quad \tilde{q}_{2n+1}(z)=\sqrt{\frac{\tau_n}{\tau_{n+1}}}q_{2n+1}^L(z) 
\end{equation}
and the recurrence relation \eqref{lskew:rec} is transformed into 
\begin{align}\label{lskew:rec2}
\begin{split}
&z(\tilde{q}_{2n+1}(z)-\alpha_{n+1} \tilde{q}_{2n}(z))=\sqrt{\beta_{n+1}}\tilde{q}_{2n+2}(z)-\tilde{q}_{2n}(z),\\
&z(\tilde{q}_{2n}(z)-\sqrt{\beta_n} \tilde{q}_{2n-2}(z))=\tilde{q}_{2n+1}(z)-\alpha_n\tilde{q}_{2n}(z),
\end{split}
\end{align}
where $\beta_N=0$ and the polynomial $\tilde{q}_{2N}(z)$ will be
\begin{equation}
\tilde{q}_{2N}(z)\propto \prod_{k=1}^N(z-z_k)(z-z_k^{-1}).
\end{equation}
The recurrence relations \eqref{lskew:rec2} can be cast in the following generalized eigenvalue problems:
\begin{align}\label{lskew:gevp}
\begin{split}
&U\bm{v}=z V\bm{v},\quad z=z_1,z_2,\ldots ,z_N,\\
&U=\begin{pmatrix}
H & I_N\\
-F^T & O
\end{pmatrix},\quad V=\begin{pmatrix}
F & O \\ G & I_N
\end{pmatrix},
\end{split}
\end{align}
where  
\begin{align}
\begin{split}
\bm{v}&=(\tilde{q}_1(z),\tilde{q}_3(z),\ldots ,\tilde{q}_{2N-1}(z),\tilde{q}_0(z),\tilde{q}_2(z),\ldots ,\tilde{q}_{2N-2}(z))^T\\
F&=\begin{pmatrix}
1 &  & & \\
-\sqrt{\beta_1} & 1 & & & \\
                & -\sqrt{\beta_2} & 1 & & \\
                &                 &\ddots & \ddots & \\
                &                 &       & -\sqrt{\beta_{N-1}} & 1 
\end{pmatrix},\\
G&=\mathrm{diag} (\alpha_1,\alpha_2,\ldots ,\alpha_N),\quad H=\mathrm{diag} (0,\alpha_1,\ldots ,\alpha_{N-1})
\end{split}
\end{align}
and $I_N$ is an identity matrix of size $N$.
It is straightforward to see that finite LSOPs appear as eigenvectors of generalized eigenvalue problem \eqref{lskew:gevp} which has a specific structure as stated below.
\begin{thm}
Matrix pencil $(U,V)$ in \eqref{lskew:gevp} is symplectic, i.e. 
\begin{align}
UJU^T=VJV^T,\quad J=\begin{pmatrix}
O & I_N \\
-I_N & O
\end{pmatrix}
\end{align}
and $V^{-1}U\in Sp(N,\mathbb{C})$.
\end{thm} 
\begin{rmk}
While the matrices $U,V$ are sparse and non-symplectic, $V^{-1}U$ is dense and symplectic. 
\end{rmk}
As we have seen the relationship between LSOPs and OPs, we immediately obtain the tridiagonal eigenvalue problem corresponding to \eqref{lskew:gevp}.
\begin{corollary}\label{cor:tridiaogonal}
Consider the tridiagonal matrix
\begin{equation}
T=\begin{pmatrix}
\alpha_1 & \sqrt{\beta_1} & & \\
\sqrt{\beta_1} & \alpha_2-\alpha_1 & \sqrt{\beta_{2}}& \\
    & \ddots & \ddots & \ddots \\
    &        & \sqrt{\beta_{N-2}} & \alpha_{N-1}-\alpha_{N-2} & \sqrt{\beta_{N-1}}\\
    &        &         & \sqrt{\beta_{N-1}} & \alpha_N-\alpha_{N-1}
\end{pmatrix}.
\end{equation}
Its eigenvalues $\{ \lambda_k\}_{k=1}^N$ are expressed in terms of the generalized eigenvalues $\{ z_k,z_k^{-1}\}_{k=1}^N$ of \eqref{lskew:gevp}:
\begin{equation}
\lambda_k = z_k+z_k^{-1},\quad k=1,2,\ldots ,N.
\end{equation}
\end{corollary} 
\section{New basis}
In the previous section, we have seen that LSOPs appear as eigenfunctions of the symplectic generalized eigenvalue problem \eqref{lskew:gevp} and also the equivalent symplectic eigenvalue problem
\begin{equation}
V^{-1}U\bm{v}=z \bm{v}.
\end{equation}
While symplectic matrix $V^{-1}U$ related to LSOPs is dense and has $2N-1$ parameters, the butterfly matrix $B$ given in \eqref{buttefly_matrix}, which is a canonical form of symplectic matrices, is sparse and has $4N-1$ parameters. 
Therefore, it is quite natural to examine whether the butterfly matrix is related with LSOPs.\par
The key observation to understand this problem is the relationship between orthogonal polynomials on the unit circle(OPUC) and CMV matrix \cite{CMV}.
While Ordinary OPUC obtained by applying Gram-Schmidt process in monomial basis $\{ 1,z,z^2,\ldots \}$ are related with some generalized eigenvalue problem, an eigenvalue problem associated with CMV five diagonal matrix can be obtained by exchanging the basis. \par
Consider the following basis
\[
\{ 1,z^{-1},z,z^{-2},z^2,\ldots \},
\]
which is similar to alternate CMV basis.
Applying Gram-Schmidt process to skew inner product $\left< \cdot | \cdot \right>_L$ in this basis, we obtain (finite) Laurent skew orthonormal Laurent-polynomials (LSOLPs) $\{ Q_{n}(z)\}_{n=0}^{2N-1}$.
In the same fashion, we find the Pfaffian expression for $\{ Q_n(z)\}_{n=0}^{2N-1 }$:
\begin{align}
\begin{split}
Q_{2n}(z)&\propto \mathrm{Pf}(0,-1,1,-2,2,\cdots ,-n,n,z),\\
Q_{2n+1}(z)&\propto \mathrm{Pf}(0,-1,1,-2,2,\cdots ,-n,-n-1,z).
\end{split}
\end{align}
From the anti-symmetry of Pfaffians \eqref{pfaff:anti2} and shift formula \eqref{lsym}, we have as a consequence
\begin{align}
Q_{n}(z)=\begin{cases}
z^{-m}\tilde{q}_{2m}(z)\quad (n=2m)\\
-z^{-m-1}\tilde{q}_{2m}(z)\quad (n=2m+1)
\end{cases}.
\end{align}
This can also be verified from the direct calculation. For instance, we have for $n\ge m$
\begin{align}
\begin{split}
\left< Q_{2m}(z)|Q_{2n+1}(z)\right>_L &= \left< z^{-m}\tilde{q}_{2m}(z)|-z^{-n-1}\tilde{q}_{2n}(z)\right>_L\\
&=\left< z^{-n-1}\tilde{q}_{2n}(z)|z^{-m}\tilde{q}_{2m}(z)\right>_L\\
&=\left< \tilde{q}_{2n}(z)|z^{n+1-m}\tilde{q}_{2m}(z)\right>_L=\delta_{mn}.
\end{split}
\end{align}
Similar calculations also show $\left< Q_{2m+1}(z)|Q_{2n+1}(z)\right>_L=\left< Q_{2m}(z)|Q_{2n}(z)\right>_L=0$.
\begin{rmk}
LSOLPs of odd degree $ Q_{2n+1}(z) $ are tantamount to even degree LSOLPs $Q_{2n}(z)$ although such correspondence does not hold for LSOPs in monomial basis $\tilde{q}_{n}(z)$.
\end{rmk}
Since we have a recurrence relation for LSOPs \eqref{lskew:rec2}, we obtain the recurrence formula for $\{ Q_n(z)\}_{n=0}^{2N-1}$.
\begin{thm}
The LSOLPs  $\{ Q_n(z)\}_{n=0}^{2N-1}$ hold the following recurrence relations:
\begin{align}\label{lsolps:rec1}
\begin{split}
zQ_{2n}(z)&=\sqrt{\beta_{n+1}}Q_{2n+2}(z)+Q_{2n+1}(z)+\alpha_nQ_{2n}(z)+\sqrt{\beta_n} Q_{2n-2}(z),\\
zQ_{2n+1}(z)&=-Q_{2n}(z),\quad n=0,1,\ldots ,N-1,\quad z=z_1,z_2,\ldots ,z_N.
\end{split}
\end{align}
\end{thm}
It seems that there are still only $2N-1$ parameters in \eqref{lsolps:rec1} and we need to add $2N$ more parameters to make a comparison with the butterfly matrix \eqref{buttefly_matrix}.
Such addition is possible since LSOLPs have two kinds of arbitrariness. One is the non-uniqueness of odd degree SOPs mentioned in \eqref{freedom1}, which comes from the definition of SOPs and still remains in case of LSOLPS.
The other is the constant multiple.
To be more specific, the map 
\begin{equation}\label{freedom2}
q_{2n}(z)\mapsto r_nq_{2n}(z),\quad q_{2n+1}(z)\mapsto r_n^{-1}q_{2n+1}(z)\quad (r_n\ne 0)
\end{equation}
does not change skew-orthogonality relation \eqref{def:sops}.
\begin{rmk}
In case of ordinary orthonormal polynomials, they also have the multiple constant freedom. However, such constant is only allowed to be $1$ or $-1$.  
\end{rmk}
Using these two arbitrary properties, we have a following theorem.
\begin{thm}
Introduce the new Laurent polynomials $\{ \tilde{Q}_{n}(z)\} _{n=0}^{2N-1}$ by
\begin{align}
\begin{split}
\tilde{Q}_{2n}(z)&=\frac{Q_{2n}(z)}{r_n},\\
\tilde{Q}_{2n+1}(z)&=r_nQ_{2n+1}(z)+\lambda_nQ_{2n}(z).
\end{split}
\end{align}
They are also LSOLPs and hold the following recurrence relations:
\begin{align}\label{lsolps:rec2}
\begin{split}
z\tilde{Q}_{2n}(z)&=\frac{\sqrt{\beta_{n+1}}r_{n+1}}{r_n}\tilde{Q}_{2n+2}(z)+\frac{1}{r_n^2}\tilde{Q}_{2n+1}(z)\\
&+(\alpha_n -\lambda_n)\tilde{Q}_{2n}(z)+\frac{\sqrt{\beta_n} r_{n-1}}{r_{n}}\tilde{Q}_{2n-2}(z),\\
z\tilde{Q}_{2n+1}(z)
&=\lambda_n\sqrt{\beta_{n+1}}r_nr_{n+1}\tilde{Q}_{2n+2}(z)+\lambda_n\tilde{Q}_{2n+1}(z)\\
&+r_n^2(\alpha_n \lambda_n -\lambda_n^2-1)\tilde{Q}_{2n}(z)+\lambda_n\sqrt{\beta_n} r_{n-1}r_n\tilde{Q}_{2n-2}(z)
\end{split}
\end{align}
for $n=0,1,\ldots ,N-1$ and $z=z_1,z_2,\ldots,z_N$.
\end{thm}
The recurrence relations \eqref{lsolps:rec2} can be cast in the following (symplectic) eigenvalue problem:
\begin{align}\label{lsolps:evp}
\begin{split}
&A\bm{w}=z\bm{w},\\
& A={\small \left(
\begin{array}{c:c}
\begin{matrix}
\lambda_0 & & &\\
         & \lambda _1 &  &\\
         &          & \ddots & \\
         &           &       & \lambda_{N-1}
\end{matrix} & \begin{matrix}
f_0 & \lambda_0e_1 & & \\
\lambda_1e_1  & f_1 & \ddots  &\\
         &       \ddots   & \ddots & \lambda_{N-2}e_{N-1}\\
         &           &    \lambda_{N-1}e_{N-1}  & f_{N-1}
\end{matrix}  \\ \hdashline 
\begin{matrix}
\frac{1}{r_0^2} & & &\\
         & \frac{1}{r_1^2} &  &\\
         &          & \ddots & \\
         &           &       & \frac{1}{r_{N-1}^2}
\end{matrix} 
  & \begin{matrix}
g_0 & \frac{\sqrt{\beta_1}r_1}{r_0} & &\\
   \frac{\sqrt{\beta_1}r_0}{r_1}  & g_1 & \ddots  &\\
         &       \ddots    & \ddots & \frac{\sqrt{\beta_{N-1}}r_{N-1}}{r_{N-2}} \\
         &           &    \frac{\sqrt{\beta_{N-1}}r_{N-2}}{r_{N-1}}  & g_{N-1}
\end{matrix}
\end{array}
\right)},
\end{split}
\end{align}
where 
\begin{align}
\begin{split}
\bm{w}&=(\tilde{Q}_1(z),\tilde{Q}_3(z),\ldots ,\tilde{Q}_{2N-1}(z),\tilde{Q}_0(z),\tilde{Q}_2(z),\ldots ,\tilde{Q}_{2N-2}(z))^T,\\
e_i&=\sqrt{\beta_i}r_{i-1}r_i, \\
f_i&=r_i^2(\alpha_i\lambda_i-\lambda_i^2-1),\\
g_i&=\alpha_i-\lambda_i,\quad i=1,2,\ldots ,N.
\end{split}
\end{align}
It should be pointed out that the matrix $A$ in \eqref{lsolps:evp} coincides with the butterfly matrix $B$ in \eqref{buttefly_matrix} if we set for $i=0,1,\ldots ,N-1$
\begin{equation}
a_{i+1}=\frac{1}{r_{i}^2},\quad b_{i+1}=\lambda_i,\quad c_{i+1}=r_{i}^2(\alpha_{i}-\lambda_i),\quad d_{i+1}=r_ir_{i+1}\sqrt{\beta_{i+1}},
\end{equation}
which is an one-to-one mapping. In the fashion similar to Corollary \ref{cor:tridiaogonal}, we can relate LSOLPS with OPs and also the butterfly matrix with tridiagonal matrix.
\begin{corollary}
Let $a_i,b_i,c_i,d_{i+1} \in \mathbb{R}$ and $a_ia_{i+1}>0$ for $i=1,2,\ldots ,N$. Consider the following tridiagonal symmetric matrix:
\begin{equation}\label{tridiaognal}
\begin{pmatrix}
a_1c_1+b_1 & \sqrt{a_1a_2}|d_2| & & & \\
\sqrt{a_1a_2}|d_2| & a_2c_2+b_2  & \ddots  & &\\ 
                   & \ddots  & \ddots & \sqrt{a_{N-1}a_N}|d_N| & \\
                   &                    &  \sqrt{a_{N-1}a_N}|d_N| & a_{N}c_N+b_N 
\end{pmatrix}.
\end{equation}
Its eigenvalues $\lambda_k,k=1,2,\ldots ,N$ are explicitly given in terms of those of the butterfly matrix:
\begin{equation}
\lambda _k=z_k+z_k^{-1}\quad k=1,2,\ldots ,N,
\end{equation}
where $z_k,z_k^{-1}$ are eigenvalues of the butterfly matrix \eqref{buttefly_matrix}. Conversely, the eigenvalues of the butterfly matrix can be obtained from those of the tridiagonal matrix \eqref{tridiaognal}:
\begin{equation}
z_k= \frac{\lambda_k+(\lambda_k^2-4)^{\frac{1}{2}}}{2},\quad z_k^{-1}= \frac{\lambda_k-(\lambda_k^2-4)^{\frac{1}{2}}}{2},\quad k=1,2,\ldots ,N.
\end{equation}
\end{corollary}
\section{Concluding Remarks}
To sum up, we have introduced a new skew inner product with Laurent symmetry which is different from one in \cite{Kodama} and examined the related SOPs, which we call LSOPs.
LSOPs hold local recurrence relations which ordinary SOPs do not. 
This can be explained by the fact that LSOPs (of even degree) are related with OPs and the new identities between Pfaffians and determinants are derived as a by-product.
Restricting the corresponding inner product space to finite space, we obtain the finite LSOPs and their recurrence relations are equivalent to a generalized eigenvalue problem, which we have shown is symplectic.
Furthermore, by exchanging the basis, we have obtained LSOLPs and they are related with some symplectic matrix.
More interestingly, this symplectic matrix coincides with (unreduced) butterfly matrix, which is a canonical symplectic matrix. 
We will offer some remarks below.  \par
Symplectic (generalized) eigenvalue problems appear in control theory and some algorithms to solve these are proposed \cite{Benner2,Bunse}.
Since butterfly matrix is canonical, the relation between butterfly matrix and tridiagonal matrix implies that the symplectic eigenvalue problems can be reduced to eigenvalue problems. 
This will indicate the new efficient algorithm for the problem.\par 
In \cite{Kodama}, it is reported that the SOPs with respect to skew inner product \eqref{inner:kodama} in finite dimension are related with the symplectic Lie algebra $\mathfrak{sp}(N,\mathbb{C})$.
It is well known that there exists a Cayley transformation, a map from symplectic matrices to symplectic Lie algebra (and vice versa):
\begin{equation}
S=(I_{2N}+s)(I_{2N}-s)^{-1}\quad (s\in \mathfrak{sp}(2N,\mathbb{C}),S\in Sp(2N,C)).
\end{equation}
This will imply the relationship among these related SOPs, which is an interesting problem for understanding SOPs and related fields more deeply. \par
As mentioned in section \ref{sec:LSOPs}, the elements of Pfaffian \eqref{moments_uc} appear in random matrix theory and symmetrized growth model. This implies the relationship between LSOPs and such models. Furthermore, in random matrix theory, (classical) SOPs are known to be derived from OPs by introducing a special operator \cite{Adler1,Forrester2} and similar techniques can be expected to introduce LSOPs.  
\section*{Acknowledgement}
The author is thankful to Satoshi Tsujimoto at Kyoto University for fruitful discussions.
He is also grateful to the two anonymous referees for their careful reading of the manuscript and helpful comments.





\bibliographystyle{elsarticle-num}



\appendix
\section{Pfaffians}
We here give a useful definition and properties of Pfaffians (For further details of elementary theory of Pfaffians, see e.g. \cite{Knuth,Ohta}). 
Pfaffians are defined by:
\begin{align}
\begin{split}
&\mathrm{Pf}(i_0,i_1,\ldots ,i_{2n-1}):=\sum_{\sigma \in \mathfrak{S}_{2n}}\frac{\textrm{sgn} (\sigma )}{n!2^n}\prod _{0\le i\le n-1}\mathrm{Pf}(i_{\sigma (2i)},i_{\sigma (2i+1)}),
\end{split}
\end{align}
where the elements of Pfaffians $\mathrm{Pf}(i,j)$ are supposed to satisfy the skew-symmetric relation:
\begin{equation}\label{pfaff:anti}
\mathrm{Pf}(i,j)=-\mathrm{Pf}(j,i).
\end{equation} 
From this definition, it is not difficult to see 
\begin{equation}\label{pfaff:anti2}
\mathrm{Pf}(\ldots ,i,\ldots,j)=-\mathrm{Pf}(\ldots ,j,\ldots ,i,\ldots ).
\end{equation}
It is useful to define Pfaffians by the following recursive procedure:
\begin{align}\label{pfaff:expansion}
\begin{split}
&\mathrm{Pf}(i_0,\ldots ,i_{2n-1})=\sum_{k=0}^{2n-2}(-1)^k\mathrm{Pf}(i_k,i_{2n-1})\cdot \mathrm{Pf}(i_0,\ldots \widehat{i_k},\cdots ,i_{2n-2}),
\end{split}
\end{align}
where $\widehat{j}$ means the deletion of $j$.
For example, 
\begin{align}
\begin{split}
\mathrm{Pf}(0,1,2,3)&=\mathrm{Pf}(0,1)\mathrm{Pf}(2,3)-\mathrm{Pf}(0,2)\mathrm{Pf}(1,3)\\
&+\mathrm{Pf}(0,3)\mathrm{Pf}(1,2),\\
\mathrm{Pf}(0,1,2,3,4,5)&=\mathrm{Pf}(0,1)\mathrm{Pf}(0,1,2,3)-\mathrm{Pf}(0,2)\mathrm{Pf}(1,3,4,5)\\
&+\mathrm{Pf}(0,3)\mathrm{Pf}(1,2,4,5)-\mathrm{Pf}(0,4)\mathrm{Pf}(1,2,3,5)\\
&+\mathrm{Pf}(0,5)\mathrm{Pf}(1,2,3,4).
\end{split}
\end{align}
Pfaffians are related to determinants in several points. One instance is that the square of Pfaffians are equalt to the determinant of skew-symmetric matrix:
\begin{equation}
\mathrm{Pf}(i_1,i_2\ldots ,i_{2n})^2=\det (\mathrm{Pf}(i_k,i_l))_{1\le k,l\le 2n}.
\end{equation}
Then it is sometimes convenient to express Pfaffians by 
\begin{equation}
\mathrm{Pf}(i_1,i_2,\ldots ,i_{2n})=\left. \begin{matrix}
| \mathrm{Pf}(i_1,i_2) & \mathrm{Pf}(i_1,i_3) & \cdots & \mathrm{Pf}(i_1,i_{2n})\\
                       & \mathrm{Pf}(i_2,i_3) & \cdots & \mathrm{Pf}(i_2,i_{2n})\\
                       &                      & \ddots & \vdots \\
                       &                      &        & \mathrm{Pf}(i_{2n-1},i_{2n}) 
\end{matrix}\right|.
\end{equation}  
In several areas including integrable systems, identities of Pfaffians play an important role. 
One of these identities is given as follows:
\begin{align}\label{pfid}
\begin{split}
&\mathrm{Pf}(M)\mathrm{Pf}(M,j_0,j_1,\ldots ,j_{2m-1})\\
&=\sum_{k=0}^{2m-2}(-1)^k\mathrm{Pf}(M,j_k,j_{2m-1})\cdot \mathrm{Pf}(M,j_0,\ldots \widehat{j_k},\cdots ,j_{2m-2}),
\end{split}
\end{align}
where $M=\{ i_1,i_2,\ldots ,i_{2n}\}$. This relation can be obtained by applying golden theorem \cite{Ohta} for Pfaffians to \eqref{pfaff:expansion}. If we take $m=2$ and $(j_0,j_1,j_2,j_3)=(a,b,c,d)$, the relation \eqref{pfid} becomes
\begin{align}\label{pfid1}
\begin{split}
&\mathrm{Pf}(i_1,i_2,\ldots ,i_{2n},a,b,c,d)\mathrm{Pf}(i_1,i_2\ldots ,i_{2n})\\
&=\mathrm{Pf}(i_1,i_2,\ldots ,i_{2n},a,b)\mathrm{Pf}(i_1,i_2,\ldots ,i_{2n},c,d)\\
&-\mathrm{Pf}(i_1,i_2,\ldots ,i_{2n},a,c)\mathrm{Pf}(i_1,i_2,\ldots ,i_{2n},b,d)\\
&+\mathrm{Pf}(i_1,i_2,\ldots ,i_{2n},a,d)\mathrm{Pf}(i_1,i_2,\ldots ,i_{2n},b,c).
\end{split}
\end{align}
Furthermore, if we take $m=3$ and $(j_0,j_1,j_2,j_3,j_4,j_5)=(i_{2n+1},i_{2n+1},a,b,c,d)$ and use \eqref{pfaff:anti2}, we have
\begin{align}\label{pfid2}
\begin{split}
&\mathrm{Pf}(i_1,i_2,\ldots ,i_{2n+1},a,b,c)\mathrm{Pf}(i_1,i_2,\ldots ,i_{2n+1},d)\\
&=\mathrm{Pf}(i_1,i_2,\ldots ,i_{2n+1},a,b,d)\mathrm{Pf}(i_1,i_2,\ldots ,i_{2n+1},c)\\
&-\mathrm{Pf}(i_1,i_2,\ldots ,i_{2n+1},a,c,d)\mathrm{Pf}(i_1,i_2,\ldots ,i_{2n+1},b)\\
&+\mathrm{Pf}(i_1,i_2,\ldots ,i_{2n+1},b,c,d)\mathrm{Pf}(i_1,i_2,\ldots ,i_{2n+1},a).
\end{split}
\end{align}
It should be mentioned that such kind of Pfaffian identieies are also known as compound Pfaffian theorem or general Wick theorem \cite{Perk}.





\end{document}